\documentclass[a4paper]{article}

\usepackage{INTERSPEECH2020}
\usepackage{algorithm} 
\usepackage{algpseudocode}
\usepackage{subcaption}
\usepackage{hyperref}

\algdef{SE}[SUBALG]{Indent}{EndIndent}{}{\algorithmicend\ }%
\algtext*{Indent}
\algtext*{EndIndent}

\title{JDI-T: Jointly trained Duration Informed Transformer for Text-To-Speech without Explicit Alignment}
\name{Dan Lim$^1$, Won Jang$^2$, Gyeonghwan O$^2$, Heayoung Park$^2$, Bongwan Kim$^2$, Jaesam Yoon$^2$}

\address{
  $^1$Kakao Corp., Seongnam, Korea\\ 
  $^2$Kakao Enterprise Corp., Seongnam, Korea
  }
\email{
    satoshi.2018@kakaocorp.com\\}

\begin{document}

\maketitle
\begin{abstract}
We propose Jointly trained Duration Informed Transformer (JDI-T), a feed-forward Transformer with a duration predictor jointly trained without explicit alignments in order to generate an acoustic feature sequence from an input text. In this work, inspired by the recent success of the duration informed networks such as FastSpeech and DurIAN, we further simplify its sequential, two-stage training pipeline to a single-stage training. Specifically, we extract the phoneme duration from the autoregressive Transformer on the fly during the joint training instead of pretraining the autoregressive model and using it as a phoneme duration extractor. To our best knowledge, it is the first implementation to jointly train the feed-forward Transformer without relying on a pre-trained phoneme duration extractor in a single training pipeline. We evaluate the effectiveness of the proposed model on the publicly available Korean Single speaker Speech (KSS) dataset compared to the baseline text-to-speech (TTS) models trained by ESPnet-TTS.
  
\end{abstract}
\noindent\textbf{Index Terms}: text-to-speech, speech synthesis, Transformer, Korean speech

\section{Introduction}
Deep learning approaches to text-to-speech (TTS) task have made significant progress in generating highly natural speech close to human quality. Especially attention-based encoder-decoder models such as Tacotron \cite{Wang2017} and Tacotron2 \cite{8461368} are dominant in this area. They generate an acoustic feature sequence, mel-spectrogram, for example, from an input text autoregressively using an attention mechanism where the attention mechanism plays the role of implicit aligner between the input sequence and the acoustic feature sequence. Finally, Griffin-Lim algorithm \cite{1172092} or a neural vocoder such as WaveNet \cite{Oord2016WaveNetAG}, WaveGlow \cite{8683143} or Parallel WaveGAN \cite{9053795} is used to convert the predicted acoustic feature sequence to corresponding audio samples.

Besides Recurrent Neural Network (RNN) based TTS models (Tacotron \cite{Wang2017}, Tacotron2 \cite{8461368}), Transformer \cite{li2019neural} has also been applied for TTS in the attention-based encoder-decoder framework successfully achieving the quality of human recording. The self-attention module, followed by a nonlinear transformation in the Transformer \cite{NIPS2017_7181}, solves the long-range dependency problem by constructing a direct path between any two inputs at different time steps and improves the training efficiency by computing the hidden states in an encoder and a decoder in parallel.

Despite its success in synthesizing high-quality speech, the attention-based encoder-decoder models are prone to the synthesis error, which prevents its commercial use. The unstable attention alignment at synthesis causes the synthesized speech to be imperfect, e.g., phoneme repeat, skip, or mispronunciation. To solve this problem, duration informed networks such as FastSpeech \cite{NIPS2019_8580} and DurIAN \cite{Yu2019DurIANDI} reduce the errors by relying on a duration predictor instead of the attention mechanism.

The reliance on the duration predictor instead of the attention mechanism is more robust since the duration predictor guarantees stepwise and monotonic alignments between a phoneme sequences and mel-spectrogram. Although the duration informed networks can synthesize high-quality speech without the synthesis error, the training process is tricky: a pre-trained model must be prepared as a phoneme duration extractor since the duration informed networks cannot be trained without a reference phoneme duration sequence.

For example, FastSpeech \cite{NIPS2019_8580} extracts a phoneme duration sequence from attention alignments matrix of pre-trained autoregressive Transformer for training a feed-forward Transformer and a duration predictor. On the other hand, DurIAN \cite{Yu2019DurIANDI} uses the forced alignment, which is commonly used in statistical parametric speech-synthesis systems to train their duration models. Thus, the previous duration informed networks have sequential, two-stage training pipeline, which may slow down the model training time. More recently, duration informed network not requiring pre-trained model \cite{9054119} has been proposed, but it still requires a multi-stage training phase.

We are motivated by a simple idea that if the attention mechanism of the autoregressive Transformer can provide reliable phoneme duration sequences from the early in joint training, the previous two-stage training pipeline of the duration informed networks could be combined. In this paper, we simplify the training pipeline of the duration informed networks by jointly training the feed-forward Transformer and the duration predictor with the autoregressive Transformer. The contributions of our work are as follow:

\begin{itemize}
    \item We propose a novel training framework where the feed-forward Transformer and the duration predictor are trained jointly with the autoregressive Transformer. By acquiring reference phoneme duration from the autoregressive Transformer during the training on the fly, the previous two-stage training pipeline of the typical duration informed networks such as FastSpeech \cite{NIPS2019_8580} or DurIAN \cite{Yu2019DurIANDI} is simplified to a single-stage training.
    \item We remedy an instability of the attention mechanism of the autoregressive Transformer by adding an auxiliary loss and adopt a forward attention mechanism \cite{8462020}. This makes the phoneme duration sequence extracted from the attention mechanism reliable from the early on in training.
    \item We prove the effectiveness of the proposed model by comparing it with popular TTS models (Tacotron2, Transformer, and FastSpeech) implemented from ESPnet-TTS \cite{9053512} on the publicly available Korean dataset.
\end{itemize}

\section{Model description}
The main idea of JDI-T is to train feed-forward Transformer and duration predictor with autoregressive Transformer jointly. In this section, we describe each component in detail.

\begin{figure}[t]
  \centering
  \includegraphics[width=\linewidth]{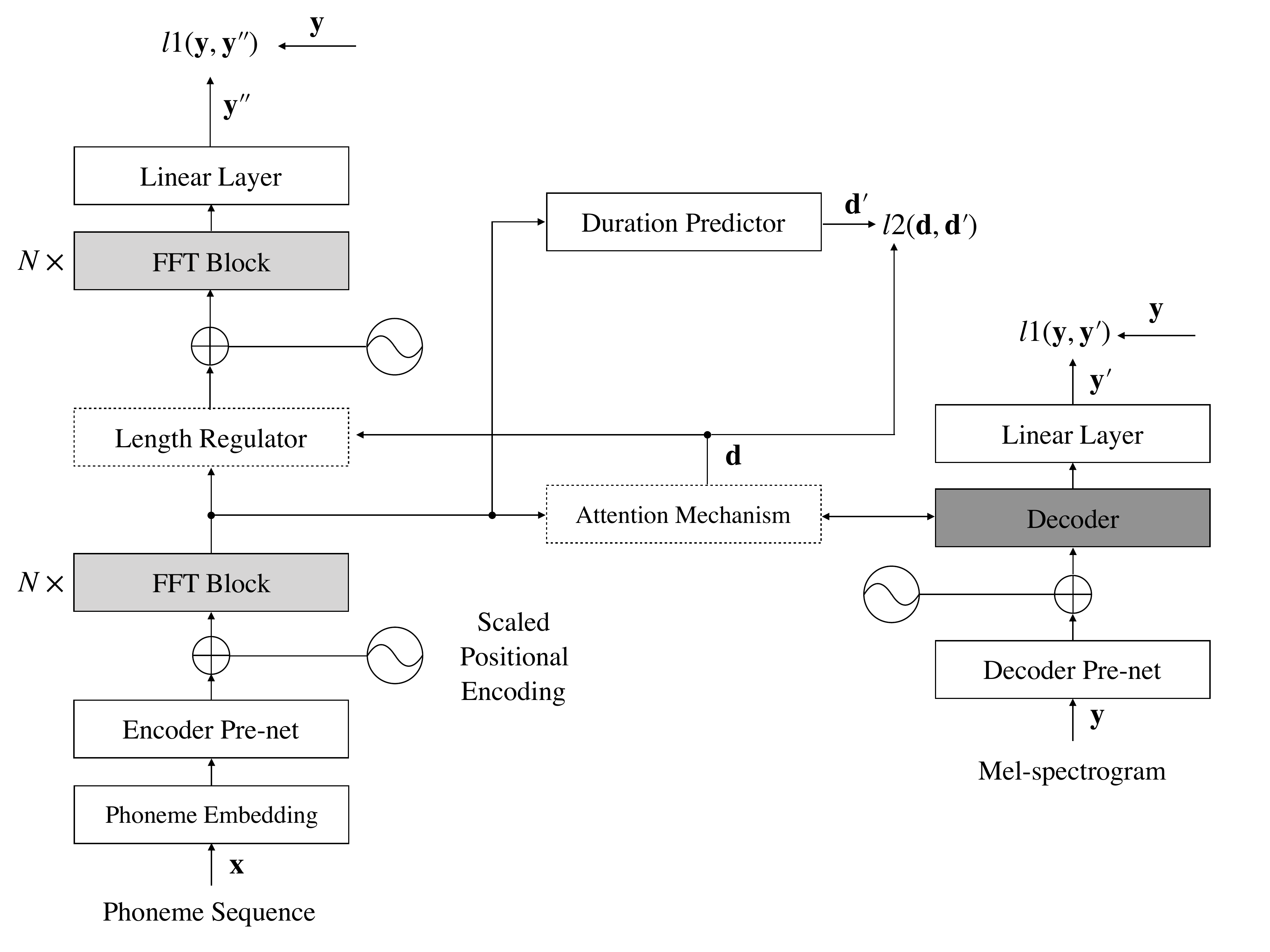}
  \caption{An illustration of our proposed joint training framework (Auxiliary loss for attention is omitted for brevity.)}
  \label{fig:fig1}
\end{figure}

\subsection{Feed-Forward Transformer}
Feed-forward Transformer located on the left in Figure \ref{fig:fig1} consists of a phoneme embedding, an encoder pre-net, scaled positional encodings, multiple Feed-Forward Transformer (FFT) blocks, a length regulator and a linear layer for phoneme sequence to mel-spectrogram transformation. 

As shown in Figure \ref{fig:sub-first}, the structure of FFT blocks is composed of a multi-head attention and a single layer 1D convolution network where residual connections, layer normalization, and dropout are used. It is slightly different from what described in \cite{NIPS2019_8580}. Note that the output of each FFT block on both sides is normalized by layer normalization. Encoder pre-net and scaled positional encoding have the same configuration as described in Transformer TTS \cite{li2019neural}. 

The stacked modules from the phoneme embedding to the FFT blocks below the length regulator works as the encoder, which is also shared by autoregressive Transformer and duration predictor. The length regulator regulates an alignment between the phoneme sequences and the mel-spectrogram in the same way described in FastSpeech \cite{NIPS2019_8580}, expanding the output sequences of FFT blocks on phoneme side according to reference phoneme duration so that total length of it matches the total length of mel-spectrogram. The feed-forward Transformer is trained to minimize $l1$ loss between predicted and reference mel-spectrogram.

\subsection{Autoregressive Transformer}
\begin{figure}[t]
\centering
\begin{subfigure}[b]{.4\linewidth}
  \centering
  \includegraphics[width=\linewidth]{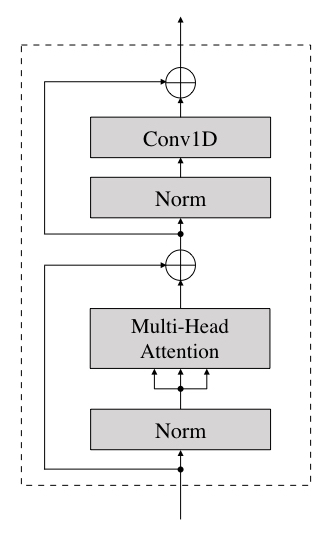}
  \caption{FFT Block}
  \label{fig:sub-first}
\end{subfigure}
\begin{subfigure}[b]{.4\linewidth}
  \centering
  \includegraphics[width=\linewidth]{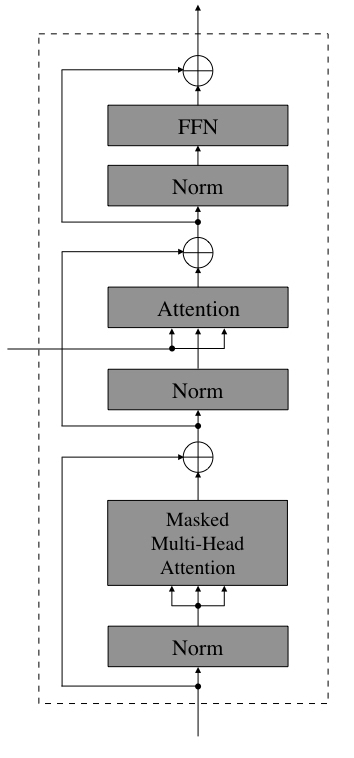}
  \caption{Decoder}
  \label{fig:sub-second}
\end{subfigure}
\caption{(a) The feed-forward Transformer block. (b) The decoder for autoregressive Transformer}
\label{fig:fig2}  
\end{figure}

Autoregressive Transformer is an attention-based encoder-decoder model as shown in Figure \ref{fig:fig1} where it shares the encoder with the feed-forward Transformer on the left bottom and the output of the encoder is attended by the stacked modules on the right consisting of a decoder pre-net, a scaled positional encoding, a decoder, and a linear layer. The decoder depicted in Figure \ref{fig:sub-second} has the similar structure with the FFT block in Figure \ref{fig:sub-first} but there are differences of having masked multi-head attention instead of multi-head attention, additional sub-layer in the middle for attention mechanism over the outputs of the encoder and position-wise Feed Forward Network (FFN) instead of 1D convolution network. Note that the output of the decoder is normalized by layer normalization. The FFN of the decoder and the decoder pre-net have the same configuration as described in Transformer TTS \cite{li2019neural}.

The autoregressive Transformer provides reference phoneme duration during training for each phoneme on the fly. Specifically, while training the autoregressive Transformer to minimize $l1$ loss between predicted and reference mel-spectrogram, phoneme duration is extracted from the attention alignment matrix by counting the number of mel frames which scores the highest value for each distinct phoneme. The phoneme duration acquired in this way is used as reference phoneme duration for the length regulator of feed-forward Transformer and the duration predictor. Note that the decoder, unlike described in Transformer TTS \cite{li2019neural}, neither has been stacked nor has multi-head attention over the outputs of the encoder since it was sufficient as a phoneme duration extractor.

\subsection{Duration Predictor}
Duration predictor consists of a 2-layer 1D convolution network with ReLU activation followed by the layer normalization, dropout layer and an extra linear layer to output a scalar value which is interpreted as phoneme duration in the logarithmic domain. The depiction of duration predictor is omitted since it is exactly same as described in FastSpeech \cite{NIPS2019_8580}. As shown in the middle top of Figure \ref{fig:fig1}, the duration predictor also shares the encoder with feed-forward, autoregressive Transformer so it is stacked on top of the FFT blocks on the phoneme side. In order to minimize the difference between predicted and reference phoneme duration, $l2$ loss is used, where the reference phoneme duration is extracted from an attention alignment matrix of autoregressive Transformer during training. Note that we stop gradients propagation from the duration predictor to the FFT blocks during joint training, since the training fails without it.

\section{Auxiliary loss and Attention mechanism}
It is very important to stabilize the attention alignments of autoregressive Transformer, since the training of the proposed model requires precise phoneme duration and the quality of phoneme duration is affected by the stability of attention alignment of autoregressive Transformer. In this section, we describe some methods for acquiring reliable phoneme duration by stabilizing attention alignments of autoregressive Transformer from the early in training steps.

\subsection{CTC recognizer loss}
In \cite{Liu2019MaximizingMI}, authors insist that the synthesis error rate of attention based TTS models can be alleviated if the model is guided to learn the dependency between the input text and the predicted acoustic feature sequences by maximizing the mutual information between them. After formulating that maximizing the mutual information is equivalent to training an auxiliary recognizer, they show that training CTC recognizer as auxiliary loss for the Tacotron TTS model can reduce the synthesis error.

Although the method of maximizing the mutual information is proposed for reducing the synthesis error of the attention-based model, we found that it also helps autoregressive Transformers learn stable attention alignments during training, which is a crucial factor for the proposed model to be jointly trained successfully. In this work, we implement it just by adding an extra CTC loss layer. Specifically, the extra linear layer is stacked on top of the decoder of the autoregressive Transformer so that it is trained to predict input phoneme sequences.

\subsection{Forward attention with Guided attention loss}
As reported in \cite{9054106}, location-relative attention mechanisms are more preferred over the content-based attention mechanisms in that they not only reduce synthesis error but also help the model align quickly during training. However, it is not proper for parallel computing. If it is used in the autoregressive Transformer such as Transformer TTS \cite{li2019neural}, the computation efficiency is sacrificed. Therefore we adopt forward attention mechanism \cite{8462020} on top of content-based attention mechanism for minimal computational overhead instead of location-relative attention mechanisms. The forward attention mechanism guarantees fast convergence and stability of the attention alignments by considering only monotonic attention alignment paths, which is a natural assumption between the input phoneme sequences and the output acoustic feature sequences. 

Algorithm \ref{alg:fa} states the forward attention mechanism adopted for the proposed model. The algorithm has same procedures as in original paper \cite{8462020} except that it prepares the attention weights $w_{1: T}(1:N)$ for all time steps $N,T$ in parallel using the attention mechanism $Attention$. The $Attention$ is a single-head, content-based attention from the decoder of autoregressive Transformer using an output sequences of the FFT blocks on phoneme side $h_{1:N}$ and an output sequences of the decoder $s_{1:T}$ which has the length of $N$ and $T$ respectively.

Although the forward attention mechanism guarantees monotonic attention alignments in principle, it is observed that the training often fails without learning any valid alignments. We found that guiding attention alignments with auxiliary loss is useful in solving the problem \cite{8461829, 8703406}. In the proposed model, the guided attention loss with the same configuration as described in \cite{8461829} is added as auxiliary loss term for training the model since it keeps a single joint training property without depending on external modules. It applies constraint on attention alignments matrix in the form of the diagonal mask based on the idea that the input phoneme sequences and the output acoustic feature sequences have nearly diagonal correspondence.

\begin{algorithm}[t]
	\caption{Forward Attention} 
	\begin{algorithmic}
	    \State \textbf{Initialize:}
	    \Indent
		\State $\hat{\alpha}_{0}(1) \leftarrow 1$
		\State $\hat{\alpha}_{0}(n) \leftarrow 0, n=2,\ldots,N$
		\State $w_{1:T}(1:N) \leftarrow Attention(h_{1:N}, s_{1:T})$
		\EndIndent
		\For{$t=1$ to $T$}
		    \State $\hat{\alpha}^{\prime}_{t}(n) \leftarrow ( \hat{\alpha}_{t-1}(n) + \hat{\alpha}_{t-1}(n-1) ) w_t(n)$
		    \State $\hat{\alpha}_{t}(n) \leftarrow \hat{\alpha}^{\prime}_{t}(n) \Big / \sum^{N}_{m=1}\hat{\alpha}^{\prime}_{t}(m)$
		    \State $c_{t} \leftarrow \sum^{N}_{n=1}\hat{\alpha}_{t}(n) h_{n}$
		\EndFor
	\end{algorithmic} 
	\label{alg:fa}
\end{algorithm}

Consequently, the loss function of the proposed model consists of two $l1$ losses for the mel-spectrogram, a $l2$ loss for phoneme duration, and two auxiliary losses, which are CTC recognizer loss and Guided Attention (GA) loss. It can be formulated as follows:
\begin{equation}
\begin{split}
    \text{L} =& \|\mathbf{y}-\mathbf{y}^{\prime}\| +
            \|\mathbf{y}-\mathbf{y}^{\prime\prime}\| + 
            \|\mathbf{d}-\mathbf{d}^{\prime}\|_2 \\
            &+ \text{CTCLoss} + \text{GALoss}
\end{split}
\end{equation}
where $\mathbf{y}$, $\mathbf{y}^{\prime}$ and $\mathbf{y}^{\prime\prime}$ are the mel-spectrograms from the reference, autoregressive Transformer and feed-forward Transformer respectively, $\mathbf{d}$ is a phoneme duration sequences from the attention mechanism of autoregressive Transformer and $\mathbf{d}^{\prime}$ is a phoneme duration sequences from the duration predictor. It can be future works to investigate an effect of different scaling constants for each loss term.

After joint training, the autoregressive Transformer is no longer needed, so only the feed-forward Transformer and the duration predictor are used for synthesis. The feed-forward Transformer generates a mel-spectrogram $\mathbf{y}^{\prime\prime}$ from the input phoneme sequences in parallel using the phoneme duration sequence predicted by the duration predictor.

\section{Experiments}
\subsection{Datasets}
We conduct experiments on two different datasets. The first dataset is an internal speech recorded by a professional Korean female speaker in studio quality. The number of utterances used for training is 25k, among which 250, 50 samples were reserved for validation and testing, respectively. The second dataset is Korean Single speaker Speech (KSS) Dataset \cite{kss2018dataset}, which is publicly available for the Korean text-to-speech task. It consists of 12,853 utterance audio files recorded by a professional female voice actress and transcription extracted from their books with a total audio length of approximately 12 hours. We reserved the last 250 utterances; 200 samples for validation and 50 samples for testing.

We convert the input text to phoneme sequences using an internal text processing tool. The acoustic feature is 80-dimensional mel-spectrogram extracted from audio with sampling rate of 22,050 Hz using the Librosa library \cite{brian_mcfee-proc-scipy-2015}. FFT size and hop size are 1024, 256 respectively. Finally, the mel-spectrogram is normalized so that every element of the feature vector has zero mean and unit variance over the training set.

\subsection{Model configuration}
The proposed model has 6 FFT blocks in the feed-forward Transformer both on the phoneme side and the mel-spectrogram side. The number of head for self-attention is set to 8, and the kernel size of 1D convolution is set to 5. The inner-layer dimension of the position-wise feed-forward network is set to 2048, and all other dimensions not stated explicitly have been set to 512. The proposed model is implemented by PyTorch \cite{NEURIPS2019_9015} neural network library and trained on 4 NVIDIA V100 GPUs with a batch size of 16 per each GPU. We optimize it using the RAdam algorithm \cite{liu2019radam} with the same learning rate schedule as in \cite{NIPS2017_7181} about 300k training steps.

For comparative experiments, we train another three models using ESPnet-TTS \cite{9053512} which is open-source speech processing toolkit supporting state-of-the-art end-to-end TTS models. The models are two autoregressive, attention-based models: Tacotron2.v3 and Transformer.v1 and a non-autoregressive, duration informed model FastSpeech.v2 following the configuration in the recipe of the toolkit. Note that the FastSpeech.v2 uses pre-trained Transformer.v1 as teacher model. 

We use Parallel WaveGAN \cite{9053795} as the vocoder for transforming the generated mel-spectrogram to audio samples in all experiments.

\subsection{Evaluation}
To evaluate the effectiveness of the proposed model, we conduct the Mean Opinion Score (MOS) test \footnote{Audio samples are available at the following URL: \url{https://imdanboy.github.io/interspeech2020}}. The proposed model, JDI-T, is compared with three different models, including Tacotron2, Transformer, and FastSpeech. The audio samples for the MOS test are generated using scripts of the test samples reserved for each dataset, and thirteen native Korean speakers listen to all of it for measuring the audio quality. Table \ref{tab:mos} shows the results on two different datasets; the Internal and the KSS. Note that the results of \textit{GT mel} are the evaluation of the audio samples converted from reference mel-spectrogram by the vocoder; thus, it indicates upper bounds that our TTS models can achieve. 

The results of the MOS test on both datasets show that the score of the duration informed model (FastSpeech) is lower than the attention-based model (Tacotron2, Transformer). Therefore it can be said that duration informed model is challenging to match the audio quality of its teacher model, especially by comparing the FastSpeech with its teacher model; Transformer. In this case, a more elaborate training technique, focus rate tuning, or sequence-level knowledge distillation, as described in \cite{NIPS2019_8580}, would be required to improve the audio quality by its teacher model. On the other hand, the score of our proposed model, which is also non-autoregressive and duration informed model like FastSpeech, is better than FastSpeech and even achieves the highest score among the TTS models in the Internal dataset. These results show that the joint training of the proposed model is beneficial for improving the audio quality as well as for simplifying the training pipeline.

In contrast, the score of the proposed model is lower than Transformer in the KSS. It may derived from the fact that the quality of audio files is poorer (i.e., the pronunciation is unclear) in KSS dataset than in the Internal dataset which is recorded in studio quality with commercial use in mind. So, it seems that the proposed model has the difficulty of learning precise attention alignments with relatively poor sound quality of audio files. It can be future works to make the model learn better alignments during joint training even in the slightly poor sound quality of audio files.

In addition to its high-quality speech synthesis, the proposed model has benefits of the robustness and fast speed at synthesis over the autoregressive, attention-based TTS models since it has the feed-forward structure and does not rely on an attention mechanism as in FastSpeech \cite{NIPS2019_8580}. Moreover, our internal test shows that Tacotron2 and Transformer have a high rate of synthesis error, especially when they are trained with the KSS dataset and synthesize the out-of-domain scripts. Note that the synthesized audio samples from the test scripts have no synthesis error.

\begin{table}[t]
  \caption{Mean opinion scores (5-point scale)}
  \label{tab:mos}
  \centering
  \begin{tabular}{ l c c}
    \toprule
    \textbf{Model} & \textbf{Internal} & \textbf{KSS} \\
    \midrule
    \textit{GT mel}              & $3.92$  & $3.87$ \\
    \textit{Tacotron2}            & $3.52$  & $3.33$ \\
    \textit{Transformer}         & $3.55$  & $3.72$ \\
    \textit{FastSpeech}          & $3.48$  & $3.23$ \\
    \hline
    \textit{JDI-T (ours)}               & $3.77$  & $3.52$ \\
    \bottomrule
  \end{tabular}
\end{table}

\section{Conclusion}
In this paper, we propose Jointly trained Duration Informed Transformer (JDI-T) for TTS. The proposed model, consisting of the feed-forward Transformer, the duration predictor, and the autoregressive Transformer, is trained jointly without explicit alignments. After joint training, only the feed-forward Transformer with the duration predictor is used for fast and robust conversion from phoneme sequences to mel-spectrogram. Experimental results on publicly available Korean datasets prove the effectiveness of the proposed model by showing that it can synthesize high-quality speech. Furthermore, it achieves state-of-the-art performance in the internal studio-quality dataset compared to other popular TTS models implemented from ESPnet-TTS.

\bibliographystyle{IEEEtran}

\bibliography{mybib}

\end{document}